\documentclass[10pt]{emulateapj}
\usepackage{apjfonts}
\usepackage{psfig}
\usepackage{epsfig}

\def\etal{et~al.}
\def\ie{{\it i.e.,\ }}
\def\eg{{\it e.g.,\ }}
\def\cf{{\it cf.,\ }}
\font\smfont=cmr9

\slugcomment{Accepted by ApJ 2005 March 10}

\begin{document}

\title{The DEEP Groth Strip Survey.  I.  The Sample\altaffilmark{1,2}}

\author{Nicole P. Vogt$^3$}

\affil{New Mexico State University, Department of Astronomy, \\
       P.O. Box 30001, Dept 4500, Las Cruces, NM 88003}
\email{nicole@nmsu.edu}

\and

\author{David C. Koo, Andrew C. Phillips, Katherine Wu$^4$, S. M. Faber, 
Christopher~N.~A.~Willmer$^{5}$, Luc Simard$^6$, Benjamin J. Weiner, Garth
D. Illingworth,
Karl Gebhardt$^7$, Caryl Gronwall$^{8}$, Rafael Guzm\'an$^{9}$, Myungshin
Im$^{10}$, \& Vicki Sarajedini$^{9}$}
\affil{University of California Observatories / Lick Observatory, Department of \\ 
Astronomy and Astrophysics, University of California,
Santa Cruz, CA 95064}

\and

\author{Edward J. Groth$^{11}$, Jason Rhodes$^{12}$, Robert Brunner$^{13}$,
Andrew Connolly$^{14}$, Alex Szalay$^{15}$, Richard Kron$^{16}$, \& Roger 
Blandford$^{17}$}
\affil{}

\begin{abstract}

The Deep Extragalactic Exploratory Probe (DEEP) is a multi-phase research
program dedicated to the study of the formation and evolution of galaxies and
of large scale structure in the distant Universe.  This paper describes the
first five-year phase, denoted DEEP1.  A series of ten DEEP1 papers will
discuss a range of scientific topics (\eg the study of photometric and
spectral properties of a general distant galaxy survey, the evolution observed
in galaxy populations of varied morphologies).  The observational basis for
these studies is the Groth Survey Strip field, a 127 square arcminute region
which has been observed with the Hubble Space Telescope in both broad I-band
and V-band optical filters and with the Low Resolution Imaging Spectrograph on
the Keck Telescopes.
Catalogs of photometric and structural parameters have been constructed for
11,547 galaxies and stars at magnitudes brighter than 29, and spectroscopy has
been conducted for a magnitude-color weighted subsample of 818 objects.
We evaluate three independent techniques for constructing an imaging catalog
for the field from the HST data, and discuss the depth and sampling of the
resultant catalogs.  The selection of the spectroscopic subsample is
discussed, and we describe the multifaceted approach taken to prioritizing
objects of interest for a variety of scientific subprograms.  A series of
Monte Carlo simulations then demonstrates that the spectroscopic subsample can
be adequately modeled as a simple function of magnitude and color cuts in the
imaging catalog.

\end{abstract}


\altaffiltext{1}{Based on observations obtained at the W. M. Keck Observatory,
which is operated jointly by the California Institute of Technology and the
University of California.}

\altaffiltext{2}{Based in part on observations with the NASA/ESA {\it Hubble
Space Telescope}, obtained at the Space Telescope Science Institute, which is
operated by AURA, Inc., under NASA contract NAS 5--26555.}

\altaffiltext{3} {University of California, Santa Cruz, CA 95064}
\altaffiltext{4} {University of Tampa, Tampa, FL 33606}
\altaffiltext{5} {On leave from Observat\'orio Nacional, Rio de Janeiro, Brazil}
\altaffiltext{6} {National Research Council of Canada, Herzberg Institute of 
Astrophysics, Victoria V9E 2E7, Canada}
\altaffiltext{7} {University of Texas, Austin, TX 78723}
\altaffiltext{8} {Pennsylvania State University, University Park, PA 16802}
\altaffiltext{9} {University of Florida, Gainesville, FL 32611}
\altaffiltext{10}{Seoul National University, Seoul, South Korea}
\altaffiltext{11}{Princeton University, Princeton, NJ 08544}
\altaffiltext{12}{Jet Propulsion Laboratory, Pasadena, CA 91109}
\altaffiltext{13}{University of Illinois, Urbana, IL 61801}
\altaffiltext{14}{University of Pittsburgh, Pittsburgh, PA 15260}
\altaffiltext{15}{John Hopkins University, Baltimore, MD 21218}
\altaffiltext{16}{Fermi National Accelerator Laboratory, Batavia, IL 60510}
\altaffiltext{17}{Stanford University, Stanford, CA 94305}

\section{Introduction}   

The advent of 8-meter class ground--based optical and infrared telescopes, the
latest generation of radio and sub-millimeter arrays, and the high spatial
resolution of the Hubble Space Telescope (HST) have produced a 100-fold
increase in observational resources with which to study the evolution of
galaxies in the local and in the distant Universe.

The commissioning of the wide--field multifiber systems used by the 2dF Galaxy
Redshift Survey (Colless \etal\ 2001) and the Sloan Digital Sky Survey
(Abazajian \etal\ 2003) have finally enabled the construction of large samples
of local galaxies, for which a wealth of detailed structural parameters have
been determined.  Parallel efforts with large optical telescopes, leveraged by
the 100-fold advantage of the latest generation of multi-object spectrographs,
are beginning to generate samples of comparable impact in the distant galaxy
field.  These will allow statistically significant analyses of the evolution
of varied populations of galaxies from the present to redshifts $z \sim 1$
(extending over half of the age of the Universe).

The Deep Extragalactic Exploratory Probe (DEEP) is a multi-phase program
focused on the study of the formation and evolution of galaxies and of large
scale structure across this redshift range.  The second phase (DEEP2),
underway at present, uses a sample of $\sim 50,000$ galaxies with
ground--based multi-band photometry and spectroscopic redshifts obtained with
the recently commissioned Deep-Imaging Multiobject Spectrograph (DEIMOS) on
the Keck 2 Telescope.  This paper describes the first five- year pilot phase
(DEEP1).  The primary observational basis is the Groth Survey Strip (GSS)
field, a 127 square arcminute region which has been observed with HST in both
broad I-band and V-band optical filters and with the Low Resolution Imaging
Spectrograph (LRIS) on the Keck Telescopes.

There have been a number of ambitious surveys bridging the gap between the
local and the distant Universe, and increasing the number counts of faint
galaxies over the last ten years (for an overview of previous work see Koo \&
Kron 1992 and Ellis 1997).  Among these works we can cite \cf the ESS survey,
Arnouts \etal\ 1997, the LDSS survey, Colless \etal\ 1999; the CNOC and CNOC2
surveys, Yee \etal\ 1996, Yee \etal\ 2000; the Hawaii Deep Fields Survey,
Cowie, Songalia, Hu, \& Cohen 1996.

Three surveys of comparable size and depth to the DEEP1 GSS survey are the
Canada-France Redshift Survey (CFRS, Lilly \etal\ 1995), the Caltech Faint
Galaxy Redshift Survey (CFGRS, Cohen \etal\ 2000), and the Very Large
Telescope Deep Survey with the VIMOS spectrograph (VLT+VIMOS, Le F\`evre
\etal\ 2004).  Figure~\ref{fig01} places the DEEP1 survey in the context
similar optical redshift surveys of the distant galaxy population, with a
direct comparison to the CFRS and CFGRS data sets.  The pioneering CFRS is a
$\sim 600$ galaxy (250 of which have HST imaging) magnitude-limited survey
which extends to $I_{AB} = 22.5$ with a median redshift of $z = 0.56$
(Crampton \etal\ 1995). It is 1.5 magnitudes shallower than the nominal limit
of DEEP1, even though the median redshifts are similar (\ie the DEEP1 survey
focuses on intrinsically fainter galaxies), due to differences in the survey
sampling algorithms and to common difficulties in obtaining redshifts for
galaxies beyond $z \sim 1.1$.  The CFGRS survey is focused upon the northern
Hubble Deep Field (HDF-N), with a comparable number of redshifts ($z_{med} =
0.7$) for galaxies extending to $R = 24$ in the HDF-N proper and to $R = 23$
in the Flanking Fields (Cohen \etal\ 2000).  The recent VLT+VIMOS survey
centers upon the southern GOODS field (Giavalisco \etal\ 2004), containing 784
redshifts for galaxies with HST imaging and an additional 815 in the
surrounding areas.  The spectral program extends down to $I_{AB} = 24$, as
does the DEEP1 survey, with a median redshift of 0.73.

Three features distinguish the DEEP GSS survey in this context.  First, the
field is a continuous band extending over $40 \times 3$ arcminutes on the sky
(covering a range of 38 co-moving Mpc by redshift $z = 1$).  Second, the
complete HST imaging in both V-band and I-band allows for the measurement of a
uniform set of rest-frame colors and structural parameters out to redshifts $z
\sim 1$.  Third, the spectral resolution of 3--4 \AA\ (versus 10 \AA\ for the
CFGRS and 40 \AA\ for the CFRS) enables both resolution of the {\sc
[O\thinspace ii]}{\rm ~$\lambda${3727} doublet feature (2.7 \AA\ rest-frame
split) and that detailed internal kinematic measurements be made for
individual objects.

\begin{figure} [htbp]
  \begin{center}\epsfig{file=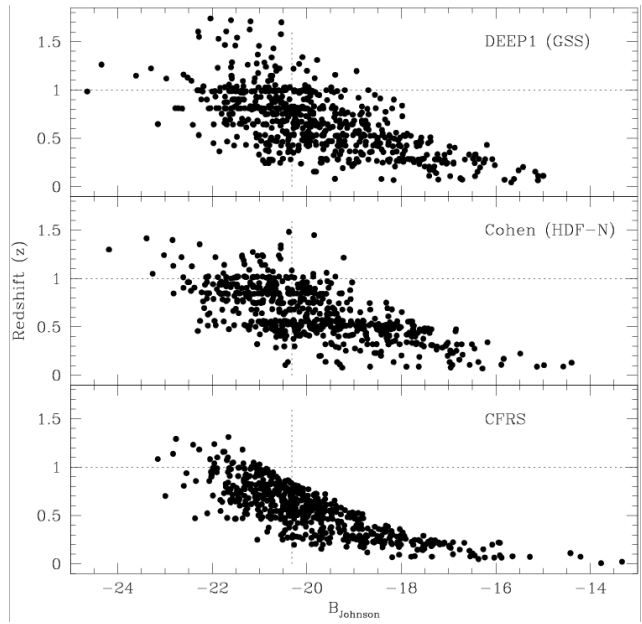, width=3.3in}\end{center}
  \caption[Figure 1] 
{The distribution of redshift $z$ versus absolute magnitude, for (a) DEEP1
(Weiner \etal\ 2005a), (b) the Caltech Faint Galaxy Redshift Survey (Cohen
\etal\ 2000), and (c) the CFRS survey (Lilly \etal\ 1995; Le F\`evre \etal\
1995; Hammer \etal\ 1995).  The dotted vertical line shows the location of
M$^{\star}$, and dotted horizontal lines mark $z = 1$ for each survey.  We
note that the lack of galaxies at redshifts well beyond $z = 1$ in DEEP1 is an
artifact of limited spectral coverage in the extreme red, rather than being
caused by a dramatic change in the underlying distribution, while the enhanced
counts at $z = 1$ are due to large scale structure across the field (\cf Le
F\`evre \etal\ 1994, Koo \etal\ 1996).  The DEEP1 survey goes $\sim 1.5$
magnitude deeper than the CFRS, comparable to the depth of Cohen \etal, and is
distinguished by complete HST+WFPC2 coverage.}
  \label{fig01}
\end{figure}

\section{Overview}

We have already published several papers which draw upon the DEEP1 GSS data
set.  Initial results of the general redshift survey were discussed in Koo
\etal\ (1996), while Vogt \etal\ (1996, 1997) found modest amounts of
evolution for disk galaxies in the field, Simard \etal\ (1999) explored the
effect of surface brightness levels in object detection from the HST images,
and Im \etal\ (2001) studied the kinematics of massive blue spheroidal
galaxies.  
This paper is the first in a formal series dedicated to these DEEP program
data.  We outline below the immediate papers published and planned within the
DEEP Groth Strip Survey Sequence:

\newcounter{bean}
\begin{list} {\Roman{bean}} {\usecounter{bean} \topsep -0.1in \itemsep -0.05in}
\item The Sample (this paper)
\item HST Structural Parameters of Galaxies in the Groth Strip (Simard \etal\ 2002)
\item Redshift Catalog and Properties of Galaxies (Weiner \etal\ 2005a)
\item Formation and Evolution of Disk Galaxies from a Sample of Spatially 
      Extended Velocity Curves (N.~P. Vogt \etal\ 2005, in preparation)
\item Evolution of Field Galaxies in Luminosity and Velocity Widths (B.~J. Weiner \etal\ 2005b, in preparation)
\item Evolution of faint AGN (V.~L. Sarajedini \etal\ 2005, in preparation)
\item The Metallicity of Field Galaxies at $0.26 < z < 0.82$ and the Evolution of 
      the Luminosity-Metallicity Relation (Kobulnicky \etal\ 2003)
\item Evolution of Luminous Bulges at High Redshift (Koo \etal\ 2005)
\item Evolution of the Fundamental Plane of Field Galaxies (Gebhardt \etal\ 2003)
\item Number Density and Luminosity Function of Field E/S0 Galaxies at $z <$ 1 (Im \etal\ 2002)
\end{list}
\vspace{0.15truein}

Updates to the our mission statement can be found on the DEEP website ({\tt
http://deep.ucolick.org}), as well as a complete data release of the GSS data.

\begin{figure*} [hbtp]
  \begin{center}\epsfig{file=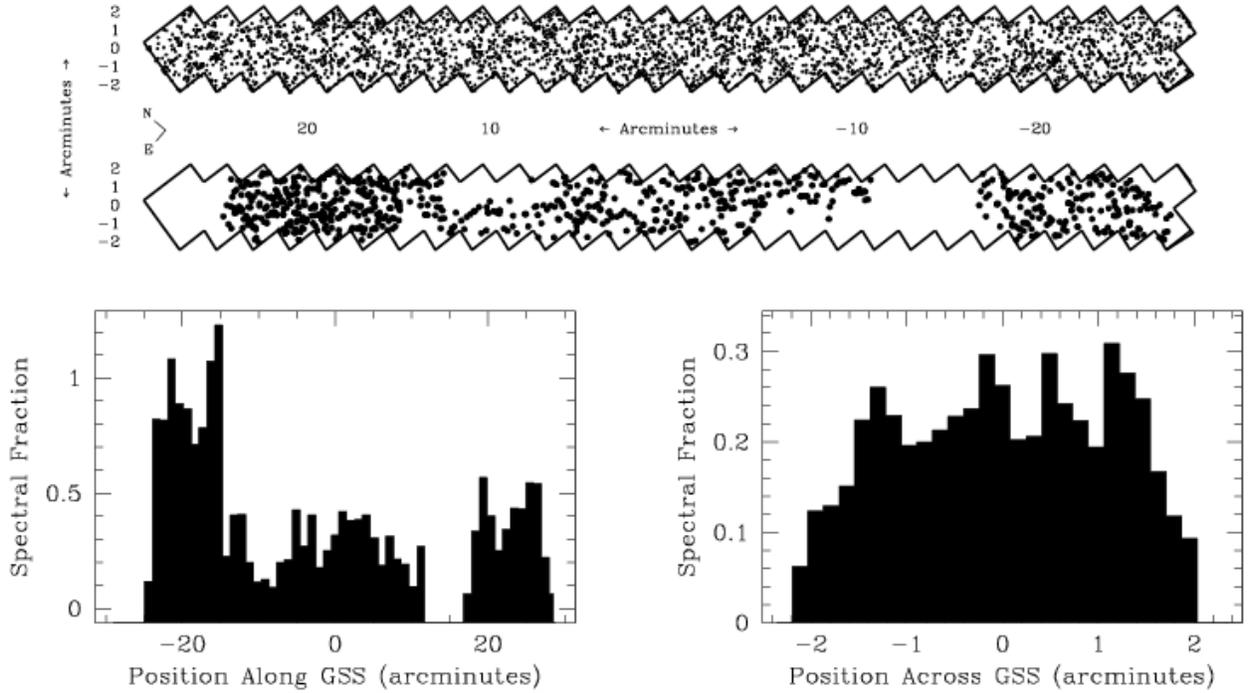, width=6.5in}\end{center}
  \caption[Figure 2]  
{The distribution of objects along the GSS.  The top two panels show the
extent of the strip on the sky, running from the first chevron (number 4, at
the northeastern extreme) to the 28$^{th}$ chevron (number 31) and covering an
area roughly forty by three arcminutes.  The upper panel shows the
distribution of all objects detected in either or both $I_{814}$ and $V_{606}$
passbands in the FOCAS catalog, and the lower panel the subset for which we
have obtained Keck+LRIS spectra.  The lower panels show the fraction of
targets within the FOCAS catalog down to the magnitude limit of our general
redshift survey for which we have spectra, as a function of position along
(left panel) and across (right panel) the GSS.}
  \label{fig02}
\end{figure*}

All Johnson-Cousins magnitudes used in this paper are on the Vega system (with
the exception of those shown in Figure~\ref{fig05}, which uses the AB system);
transformations between the Vega and AB systems are listed in Appendix~B.  We
have chosen to use a $\Lambda$CDM cosmogony with $h=0.7, \Omega_M=0.3$, and
$\Omega_\Lambda=0.7$ for this entire series of papers, unless otherwise noted.

\section{Field History}

The GSS field is a high latitude field in the HST continuous viewing zone with
no bright stars and with marginal levels of galactic extinction.  The primary
science drivers for the initial HST+WFPC2 Cycle 4 observations were to study
cosmology and galaxy evolution by evaluating the evolution of number counts,
the luminosity function, colors and morphology, and the distribution of
galaxies and of clusters in redshift.

The field is defined as the spatial position of a long band of imaging data
from two Hubble Space Telescope and Wide Field Planetary Camera (HST+WFPC2)
programs (GTO 5090 with PI E. Groth; GTO 5109 with PI J. Westphal).  They
consist of 28 overlapping WFPC2 chevrons, numbered 4 through 31 because three
chevrons originally placed at the northeast end of the strip were discovered
to lie too far from appropriate guide stars and were replaced with three new
pointings at the south-west end.  The field is oriented at $45^{\circ}$
eastwards of north, centered at ($14^h17^m, \, +52^{\circ}$), as shown in
Figure~\ref{fig02}, has a total area of 127 square arcminutes, and lies at a
Galactic latitude of $b \sim 60^{\circ}$.

Though the strip of WFPC2 chevrons runs along an angle of $45^{\circ}$
eastwards of north, the orientation of the roll angle, equivalent to the axis
of symmetry of each chevron along the line bisecting each chip 3 from pixel
(0,0) to (800,800), lies at $40.5^{\circ}$ eastwards of north.  Because of the
$4.5^{\circ}$ offset, chips 2 of each chevron overlap slightly within a region
of size $79^{\prime\prime}$ by $21^{\prime\prime}$ with the chips 3 of the
next chevron along the strip.  These overlap regions were used to align the
chevrons together in to a single supermosaic for the data taken within each
passband, as were diffraction spikes that crossed chip to chip boundaries.

Each chevron has been imaged for 2,800 seconds in a broad $V$ filter (F606W,
$V_{606}$) and for 4,400 seconds in a broad $I$ filter (F814W, $I_{814}$),
with the exception of the single deep chevron (number 7, fourth from the
northeast edge, centered at $14^h17^m30^s, \, +52^{\circ}30^m$) which was
observed for 24,400 seconds in $V_{606}$ and for 25,200 seconds in $I_{814}$.
Images for the primary program were constructed by rejecting cosmic rays from
four undithered exposures in each filter, each lasting 700 seconds in $V$ or
1100 seconds in $I$.  (A series of four orbits were spent observing each
chevron, with each orbit consisting of a single exposure in $V$ and a single
exposure in $I$.)  For the deeply imaged chevron, eight 2100 second and four
1900 second exposures, each taking a full orbit, were combined in $V$, and
twelve 2100 second exposures were combined in $I$.

Additional WFPC2 optical imaging data have been collected along the GSS by a
variety of authors and teams, including $3 \times 2,800$ seconds in a broad
$B$ filter (F450W, $B_{450}$) and $3 \times 2,800$ seconds in $I_{814}$ (GO
5449 with PI S. Lilly) for a chevron which overlaps by 70\% with the third GSS
chevron (number 6).  In Cycle 9, all GSS chevrons but the deeply imaged number
7 were re-imaged for $700$ seconds with $V_{606}$ (GO 8698 with PI J. Mould).  
A small number of objects within the Groth Strip have been selected for
additional observations with other HST instruments (NICMOS GO 7871 with PI
A. Connolly; NICMOS GO 7883, STIS GO 10249 with PI N. Vogt).

Keck+LRIS ground-based BRI imaging of the northeast one-third portion of the
Groth Strip has also been conducted by the DEEP team.  Shallower UBRI imaging
(down to U = 25, B = 25.5, R = 23.4, I = 23.8 for S/N = 10) of the same region
was done at KPNO, and may be extended as a part of the DEEP2 survey.
Additional data have been obtained at longer wavelengths for parts of the
field using SCUBA and the VLA (Fomalont \etal\ 1991), as well fairly extensive
data in K-band (\cf Crist\'obal-Hornillos \etal\ 2003; Conselice \etal\ 2005).
The field also has been or will be observed as a part of numerous large survey
programs (\eg the CFHT Legacy Survey; also with XMM, Chandra, Galex, JWST, and
ISO).  This list is not a complete census; we simply wish to illustrate that a
significant amount of observational resources, spanning multiple wavelength
regimes, have been devoted to this portion of the sky.

\section{Astrometric Solution}

An initial astrometric and photometric analysis of the GSS WFPC2 data was
completed with the Faint Object Classification and Analysis System (FOCAS;
Jarvis \& Tyson, 1981) for 11,547 objects down to 33$^{rd}$ magnitude in
$(V_{606} + I_{814})/2$.

An independent astrometric solution was created for a supermosaic in each
passband composed of all 28 chevrons, by first transforming the individual
chip coordinates to a common coordinate system based on the polynomial
coordinates derived by Holtzman et al. 1994.  The rms residual errors in $x$
and in $y$ on each chip are equal or less than $0.02^{\prime\prime}$ along the
entire strip, as determined by evaluation of the objects observed on two
chevrons in the overlap regions.  The values are fairly uniform, though we
note that the residuals on the chip 3 regions are slightly higher on average
(by 25\%) than those for chips 2 and 4.

There are six HST guide star catalog objects within the strip, which were used
for astrometric zero pointing.  The error in guide star coordinates is a
sizable fraction of an arcsecond (we estimate 0.5$^{\prime\prime}$ in each
coordinate), which will affect the absolute calibration.  Star positions were
measured on the WFPC2 chevrons by measuring the intersections of diffraction
spikes; in one case the stellar flux peak could also be measured, and the two
techniques agreed to within a pixel.  An rms residual of $0.33^{\prime\prime}$
was obtained by using all six stars, and the best, final fit was obtained from
four stars (rms value of $0.11^{\prime\prime}$), by discarding two stars which
fell very close to the edge of the WFPC2 chips.

Because of their late attachment to the Cycle 4 observational program, we note
that chevrons 28 - 31 were taken with an offset in roll angle of
0.6004$^{\circ}$ relative to the remainder of the strip (the roll angle was
specified to only 0.01$^{\circ}$ in the original observing proposal).  In
contrast, the average error in the roll angle was measured to be 0.2$^{\circ}$
across the entire Cycle 4 WFPC2 data set.  These factors were taken into
account in the astrometric solutions.  A comparison of the final solutions for
the $V_{606}$ and $I_{814}$ data yielded errors of less than
$0.01^{\prime\prime}$ between the two passbands; this close agreement allowed
us to overlay the two mosaics on top of each other to produce a valid color
map of the entire strip.

Summarizing the primary sources of error in the astrometric model, small scale
(\ie chevron scale) errors are controlled by the chip distortion model
($0.02^{\prime\prime}$ rms), chevron offset determination
($0.025^{\prime\prime}$ rms), and roll errors ($0.025^{\prime\prime}$ rms in
the chip corners furthest from the chevron center), leading to a total rms
error of $0.035^{\prime\prime}$.
This translates into rms errors of order $0.1 ^{\prime\prime}$ in right
ascension and declination, and a rotational uncertainty of 0.017$^{\circ}$,
along the entire strip.
When evaluating individual objects, one must of course also take into account
the errors in object centroiding (discussed below).

\begin{table*} [hbtp]
  {\small
  \caption{Catalog of Aperture Photometry}
  \begin{center}
  \begin{tabular} {l l l l r@{:}r@{:}l r@{:}r@{:}l r r r r r r r l} 
  \hline
  \hline
  \multicolumn{4}{c}{Identification Codes} & 
  \multicolumn{6}{c}{Coordinates (J2000)}  & 
  \multicolumn{7}{c}{Magnitudes$^d$}       & 
  Selection                               \\ 
  GSS & G$^a$ & B$^b$ & CFRS$^c$                      &
  \multicolumn{3}{c}{R.A.}     & \multicolumn{3}{c}{Dec.}    &
  \multicolumn{1}{c}{$V_{c}$}  & \multicolumn{2}{c}{$~I_{c} ~~V_{c\prime}-I_{c\prime}$} & 
  \multicolumn{1}{c}{$V_{f}$}  & \multicolumn{1}{c}{$I_{f}$} & 
  \multicolumn{1}{c}{$V_{g}$}  & \multicolumn{1}{c}{$I_{g}$} & 
  Criteria \\
  \hline
  072\_4040 & G82a  & B08839 &          & 14 & 17 & 37.836 & +52 & 28 & 58.45 & 22.15 & 20.92 & 1.27 & 20.64 & 19.63 & 19.58 & 20.61 & disk     \\
  073\_0542 & G1307 & B09464 &          & 14 & 17 & 46.405 & +52 & 28 & 39.65 & 24.70 & 23.63 & 1.08 & 24.59 & 23.55 & 22.06 & 23.43 & zsurvey  \\
  073\_1810 & G1318 & B09321 &          & 14 & 17 & 42.635 & +52 & 28 & 45.32 & 23.48 & 22.01 & 1.43 & 21.96 & 20.54 & 21.30 & 22.85 & morph    \\
  073\_2356 & G1355 & B09831 &          & 14 & 17 & 47.401 & +52 & 29 & 00.50 & 23.51 & 22.89 & 0.64 & 23.42 & 22.82 & 22.55 & 23.09 & serendip \\
  073\_3539 & G1367 & B09751 & C14.1043 & 14 & 17 & 45.385 & +52 & 29 & 08.27 & 22.04 & 20.35 & 1.77 & 21.19 & 19.65 & 19.60 & 21.12 & phz      \\
  \hline \\ [-0.1in]
  \multispan{14} $^a$See Rhodes, Refregier, \& Groth (2000);  $^b$see Brunner, Connolly, \& Szalay (1999);  $^c$see Lilly et al. (1995);  $^d$HST F606W \hidewidth \\
  \multispan{14} and F814W magnitudes, $V_{606}$ and $I_{814}$ within 1.\arcsec 5 diameter, $V_{606} - I_{814}$ within 1\arcsec\ diameter;  $V_{606}$ and $I_{814}$ measured with  \hidewidth \\
  \multispan{14} FOCAS; $V_{606}$ and $I_{814}$ measured with GIM2D.                                                \hidewidth \\
  \end{tabular}
  \end{center}
  \label{tab01}
  }
\end{table*}

\section{Photometric Catalogs}

The photometric catalog was created using FOCAS within the Image Reduction and
Analysis
\footnote{IRAF is distributed by the National Optical Astronomy
Observatories, which are operated by the Association of Universities for
Research in Astronomy (AURA) under cooperative agreement with the National
Science Foundation.}
(IRAF) environment to determine valid detections, classified as 10,955
galaxies, 436 stars, and 156 objects of uncertain morphology (\eg fuzzy,
diffuse) where classification was done by rescaling (broadening and narrowing)
the point spread function (PSF) to create a set of templates then used to best
match each object's flux distribution.  The objects identified as stars span
the full range of magnitude of the sample, and represent a conservative
estimate of the stellar contribution.  The spectral subsample contains 37
objects classified as stars; divided between 28 objects at $z = 0$ and nine
for which a redshift could not be determined (due to \eg faint flux, lack of
distinguishing features).  An additional 128 spectra were obtained for stars
which had been misidentified as galaxies by FOCAS.  In summary, FOCAS
successfully identified from 55 to 73\% of all stars brighter than 19$^{th}$
magnitude, dropping to the range of 10 to 22\% for all magnitudes, within the
spectroscopic subsample.  There was in general great difficulty in
distinguishing between intermediate redshift ($z \sim 0.7$) compact galaxies
and stars, \eg halo M dwarfs, solely from photometric data.  Note that the
FOCAS object classification codes were not used during the process of
selecting spectroscopic subsamples.

FOCAS operates by detecting objects of a set minimum size that register above
a set threshold level above a sky background level.  Background levels were
determined for each chip (2 through 4, note that we did not use the planetary
camera chip data, so as to maintain a uniform resolution and sensitivity
throughout) and on each chevron, by examining blank rectangles of sky.  The
detection threshold was set to $3\sigma$ above sky background for the bulk of
the GSS, and to $4.5\sigma$ for the deeply imaged field (chevron 7).  When
making this evaluation, the minimum object size was set to two pixels
($0.2^{\prime\prime}$) across the entire strip.

Three objects were manually added to the catalog of FOCAS detections, each
being a target that fell well within the extended flux of a brighter galaxy or
stellar diffraction spike and thus was not recognized as a separate object,
but for which we wished to obtain a redshift nonetheless.

After each object detection, FOCAS computes an individual sky background
surrounding each object.  Each object was then ``grown'' by adding concentric
rings of pixels until its total area had doubled.  The grown object was then
evaluated to determine its magnitude and centroid.  FOCAS attempts to split
blended objects automatically, but the algorithm used for this purpose was not
found to be optimized for faint object detections on these images (\ie down to
($V_{606} + I_{814})/2 = 25$ completeness limit, as discussed below).  It was
thus necessary to examine each object by eye, to verify and on occasion to
correct the division of parent objects in order to produce the proper
splitting into de-blended objects.  The low detection threshold and small
object size also introduced spurious ``noise'' objects, which were filtered
out by matching $V_{606}$ and $I_{814}$ frame detections.

One potential drawback to this technique is that the size of the detection
area for a given object can vary between $V_{606}$ and $I_{814}$ frames, as
the data for the two bandpasses have different levels of sensitivity.  In
addition, an actual variation in object size can be a function of $V_{606} -
I_{814}$ color, one of the key catalog parameters which was used in selecting
a subsample for follow-up spectroscopic observations.  Object centroiding was
also done independently in each bandpass, leading to shifts in object position
between the two bandpasses.  Taken together, these factors might strongly
affect the applicability of the ``total'' magnitudes derived from FOCAS for
our fainter candidates.

We note that the $V_{606}$ catalog data appear to extend more deeply than
those of the $I_{814}$ observations (\ie more objects are detected in
$V_{606}$ but not in $I_{814}$ than the reverse).  This effect becomes
significant for objects fainter than 25 in $V_{606}$, beyond the limits of the
spectral sample.  Down to 25 in $V_{606}$ and to 24 in $I_{814}$ (\ie
extrapolating a median $V_{606} - I_{814}$ color of one for non-detections),
there are 171 objects detected only in $V_{606}$ and 157 more detected only in
$I_{814}$ (and within the spectral sample only 16 targets were undetected in
$I_{814}$ and 23 were undetected in $V_{606}$).  In contrast, when we extend
these limits by one magnitude we find 1681 objects detected only in $V_{606}$
versus 335 detected only in $I_{814}$, and the numbers increase to 5198 versus
760 when the entire optical catalog is evaluated.

This increase in faint $V_{606}$ detections is caused by several factors.
First, object detection was set to penetrate well into the noise, and the
faint detection threshold was set independently in each bandpass and extends
to fainter levels in $V_{606}$.  Second, the narrower $V_{606}$ point spread
function (\cf Casertano \etal\ 2000) allowed ``blobby'' objects to be split
into multiple components, while the same group of objects (or, alternatively,
the same galaxy plus {\rm H} {\smfont II} regions) was treated as a single
object with the summed, brighter magnitude in the $I_{814}$ data.

We defined a set of object names according to the following convention.  Every
object within the imaging catalog received a seven-digit name, optionally
ended with a single letter.  The first three digits are the WFPC2 chevron
number along the strip (two digits between 4 and 31, with a leading zero if
necessary) and chip (single digit, one of chips 2 through 4); they are
followed by an underscore symbol. The final four digits are the x and y
position of the object on its chip as detected in $I_{814}$, if so detected
(else in $V_{606}$), rounded off to the nearest ten pixels. For example, an
object found in field 7 (the deeply imaged field) on chip 4 at coordinates
(235, 516) would thus be named 074\_2452.

One can conceivably (though rarely) have multiple objects with the same name
(when objects are found within the same 1" square on a chip), so we also track
the FOCAS-generated identification number (running from 1 to 11,543) to avoid
any confusion.  When we do include two such objects in the spectral sample,
the second to be observed has an "a" attached to the end of its seven-digit
name.

We elected to conduct a second photometric analysis of the data based on
aperture magnitudes, measured with the IRAF package APPHOT.  Object centroids
as determined from the FOCAS $I_{814}$ data were used to define the location
of all targets within both bandpasses, supplemented by the $V_{606}$ object
centroids for those undetected in $I_{814}$.  Aperture magnitudes were
determined for both $1.0^{\prime\prime}$ and $1.5^{\prime\prime}$ diameter
regions for each object detected by FOCAS.

We analyzed 421 objects brighter than 25$^{th}$ magnitude within the overlap
regions of successive chips 2 and 3 twice (\ie once per separately observed
chip, or chevron) to determine rms error rates.  Object centroids had been fit
to the chip 3 data with the FOCAS catalog; these were used to determine object
positions on the overlapping chip 2 regions.  The offset between successive
chevrons varied by $\pm 2^{\prime\prime}$ across the entire GSS, and was
determined for each chevron by examining the most compact objects in the
overlap region.  No rotation correction was applied, as the linear offsets
were deemed sufficient.  The individual derived chip 2 object positions were
then allowed to shift to match the local maximum, resulting in sub-pixel
corrections.  Background levels were estimated on a chip-by-chip basis, and
then aperture photometry was performed using set $1.0^{\prime\prime}$ and
$1.5^{\prime\prime}$ diameter apertures for both sets of images.

The differences in flux levels were found to be fairly constant, at $0.025 \pm
0.030$ magnitudes throughout the $I_{814}$ data set.  Comparison values for
the $1.0^{\prime\prime}$ and $1.5^{\prime\prime}$ apertures differed by
$0.003$ magnitudes, and there was no significant trend with magnitude for
objects brighter than 25$^{th}$ magnitude.  This is likely a combination of
two competing effects.  The brighter objects exhibit high flux levels but
extend spatially beyond the aperture limits, and so small variations in
centroiding between chips 2 and 3 can cause variations in estimating the
aperture flux.  The fainter objects are contained almost entirely within the
apertures, and thus centroiding issues become less relevant, but the overall
signal to noise levels decrease.  The $V_{606}$ data showed slightly larger
differences between chip 2 and chip 3 measurements, at $0.035 \pm 0.035$
magnitudes, due in part to slightly elevated background levels.

For spectroscopic target selection, we elected to use the $V_{606}$ and
$I_{814}$ magnitudes as measured in the $1.5^{\prime\prime}$ diameter
apertures, combined with the $1.0^{\prime\prime}$ diameter aperture V$_{606}
-$ $I_{814}$ colors (``Core'' colors) to define object magnitudes and colors.
This improved the accuracy of color measurements (\ie a relative flux
measurement), particularly for faint detections, and permitted us to extend
color and two-bandpass measurements to faint objects that had been detected by
FOCAS in only one of the two HST bandpasses.  We label this catalog of
magnitudes and colors the ``Core'' catalog, as the object fluxes are sampled
only in the core regions (\ie aperture size is constant, and does not vary
with object angular size).

Table~\ref{tab01} shows a representative set of entries to the object catalog,
available in its entirety on the DEEP web site at {\tt
http://deep.ucolick.org}.  Each object is identified with its DEEP1 seven to
eight element name, followed by the FOCAS catalog identification and the names
(as available) in the ground-based UBRI (Brunner, Connolly, \& Szalay 1999)
and in the CFRS survey object catalogs.  Coordinates are listed (J2000),
followed by aperture $V_{606}$ and $I_{814}$ aperture magnitudes and V$_{606}
-$ $I_{814}$ colors.  The final column indicates the selection criteria by
which the object was placed on a Keck+LRIS mask, if this occurred (see
discussion of keywords below).

A third photometric analysis was conducted with the galaxy modeling package
GIM2D after the majority of the spectral program had been completed, to derive
a detailed, extended set of structural and morphological parameters for all
objects.  Discussion of this ``GIM2D'' catalog, and its optimization, is the
primary topic of Paper II (Simard et al. 2002).  These data, including
magnitudes as well as structural parameters, are used throughout most of the
DEEP1 science papers.

\begin{figure} [htbp]
  \begin{center}\epsfig{file=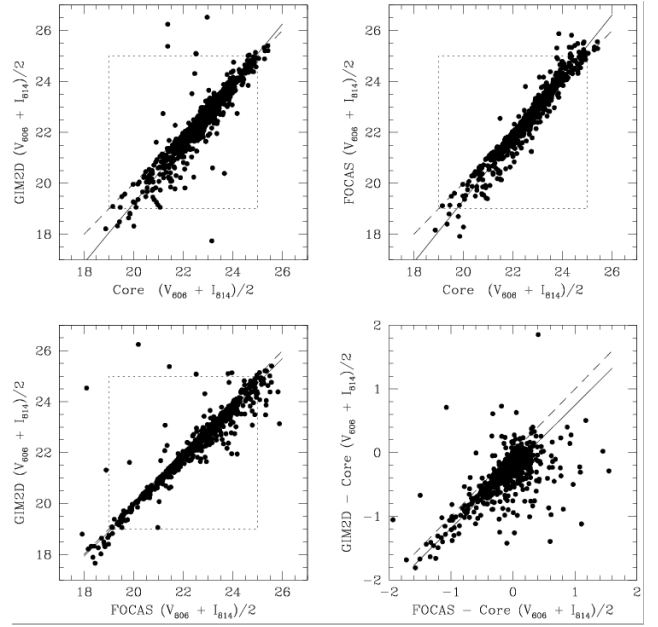, width=3.3in}\end{center}
  \caption[Figure 3]  
{The dispersion of magnitudes throughout the three imaging analyses (FOCAS
total, $1.5^{\prime\prime}$ aperture Core, and GIM2D total) used for the GSS
sample.  The first three panels show the relative distribution of magnitudes
between catalogs; the dashed line follows the relation $y = x$ and the solid
line is an unweighted fit to objects between 19 and 25 magnitudes (boxed)
differing by less than 2 magnitudes between catalogs.  The final panel plots
the difference between FOCAS and Core magnitudes versus the difference between
GIM2D and Core magnitudes, with a fit to the same points as in the other
panels.  We find that the Core magnitudes systematically underestimate
luminosity for the brightest objects, because their large angular diameters
extend beyond the $1.5^{\prime\prime}$ Core apertures, while the difference
between FOCAS and GIM2D total magnitudes becomes more significant for objects
fainter than 23$^{rd}$ magnitude.  The final panel demonstrates that the
offset for the Core aperture measurements is more significant than the scatter
between the two estimates based on total magnitudes, as the data tend to be
forcibly extended along the $y = x$ line by the Core offsets (with the
brightest objects extending this trend preferentially to the lower left hand
quadrant) rather than scattering uniformly about the origin.}
  \label{fig03}
\end{figure}

\begin{figure} [htbp]
  \begin{center}\epsfig{file=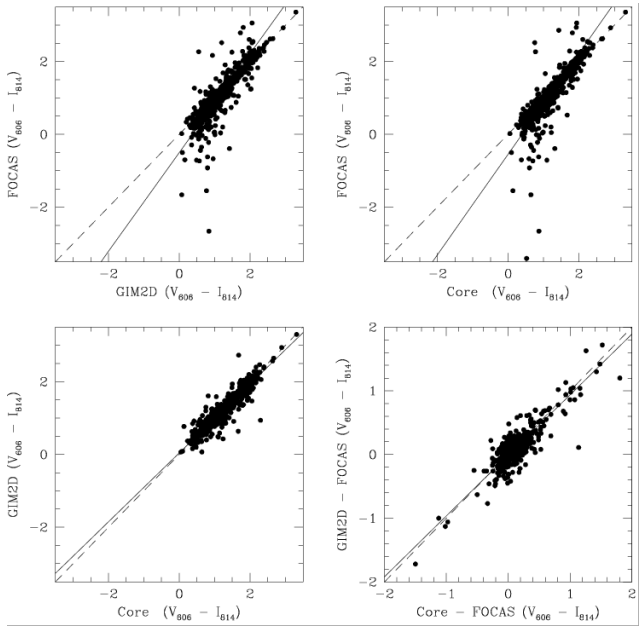, width=3.3in}\end{center}
  \caption[Figure 4]  
{The distribution of colors throughout the three imaging analyses (FOCAS
total, GIM2D total, and $1.0^{\prime\prime}$ aperture Core) used for the GSS
sample.  The first three panels show the relative distribution of V$_{606} -$
$I_{814}$ color between catalogs; the dashed line traces the relation $y = x$
and the solid line is an unweighted fit to objects between 19 and 25
magnitudes differing by less than 2 magnitudes between catalogs (no
significant change appears in the fits if we relax these magnitude limits).
The final panel plots the difference between GIM2D and FOCAS colors versus the
difference between Core and FOCAS colors, with a fit to the same points as in
the other panels.  The FOCAS-derived colors contain some substantial outliers,
particularly in a population of objects with extremely blue measurements of
color which are not reproduced in the other two catalogs.  The final panel
demonstrates that this offset is more significant than the difference between
the GIM2D and Core estimates of colors (which are offset by less than half a
magnitude), as the data tend to be forcibly extended along the $y = x$ line by
the FOCAS offsets rather than scattering uniformly about the origin.}
  \label{fig04}
\end{figure}

Figures~\ref{fig03} and \ref{fig04} demonstrate the relation between
photometric measurements made through these three analyses.  We emphasize
that, though the photometric analyses are independent for each catalog, the
object centroids were all drawn from the FOCAS catalog and were not relocated
in subsequent analyses (\ie we have preserved the link between an given object
name and its position of the WFPC2 chevrons, within all three catalogs).
The ``Core'' magnitudes are seen to systematically underestimate the fluxes of
the brightest objects, as expected given that their angular sizes extend well
beyond the $1.5^{\prime\prime}$ diameter aperture.  Note that the effect is
most significant for objects brighter than 23$^{rd}$ magnitude, which were all
placed in a single magnitude bin during our spectral target selection process,
and does not vary significantly with $V - I$ color.  When comparing the two
independent estimates for total magnitude, we find that the dominant source of
variation is the scatter in the measurements at faint (below 23$^{rd}$)
magnitudes.

There are a few ($N \sim 20$) objects with large offsets from catalog to
catalog; these extreme differences, on the level of two to five magnitudes,
are primarily caused, in order of frequency and importance, by (a)
difficulties in measuring fluxes for objects on the extreme edges of chips
(centered within the outer 7$^{\prime \prime}$, where the background
measurement has been compromised by the sharp decline in pixel responsivity
and differences in estimating background levels assume a critical role), (b)
overlapping objects which are not deconvolved identically by each set of
algorithms, and (c) stellar diffraction spikes.  We examined all such
outliers, and found only one which could not be explained by one of these
three factors: 063\_2764, an extremely diffuse, extended low surface
brightness object.

The FOCAS and Core catalogs show the fewest such offsets, as shown in the
first three panels of Figure~\ref{fig03}.  The GIM2D catalog is more sensitive
to the first, and most common, cause of this type of error because its
algorithms utilize a larger region around each object when determining the
background flux level.  It is thus likely that the GIM2D magnitudes are the
least accurate measure of flux for an object when a substantial (greater than
two magnitudes) difference exists between catalog entries.

When examining relative $V_{606}$ and $I_{814}$ magnitudes (\ie colors), we
found that the independent centroiding between passbands in the FOCAS
measurements produced a significant scatter in color overall, and an increase
in colors at the extremes of the color distribution (particularly on the blue
end).  We describe two representative cases for clarity.  
In the first case FOCAS identified two objects separated by
0.9$^{\prime\prime}$ on the sky in $I_{814}$, one of magnitude 21.8 and the
other 25.3.  In $V_{606}$, a single object with magnitude 22.6 was placed
within the same region.  The single bright $V_{606}$ detection was then
matched to the fainter $I_{814}$ object, while the brighter $I_{814}$ object
was assumed to have been undetected in $V_{606}$.  The resultant mismatch
produced an object with an artificially low $I_{814}$ flux, resulting in an
extremely blue color index.
In the second case FOCAS split a ``blobby'' object into three separate objects
on the $V_{606}$ image, while melding the total flux into a single object on
the $I_{814}$ frame.  The match of the total object in $I_{814}$ versus a
single component in $V_{606}$ produced an extremely red color index.  These
outlying cases can be characterized by total object areas which differ greatly
(by factors of 10 to 90) between the two frames (versus a factor of $\sim$
three across the rest of the data set).
These objects are mainly found in the luminosity range between 23$^{rd}$ and
25$^{th}$ magnitudes.  Extremely blue and extremely red objects were rated to
be of considerable importance within our spectroscopic program, thus it was a
priority to determine accurate colors for them.

The use of a single set of object positions for both the $V_{606}$ and the
$I_{814}$ images in the Core and GIM2D catalogs provides a more robust
estimation of the relative flux (though the background levels were still set
independently within each passband).  We note that the 1.0$^{\prime\prime}$
aperture Core colors are $0.03 \pm 0.15$ magnitudes redder than those
determined for the entire galaxy with the GIM2D package, where the positive
offset could be due to a slight nuclear concentration of redder stellar
populations in many objects.

One of the science projects driving the spectral program was the modeling of
photometric redshift estimates based on multi-color broad-band photometry
(Brunner 1997; Brunner, Connolly, \& Szalay 1999).  We were actively searching
for a small population of faint galaxy candidates with extremely blue colors,
to use to improve the models, as well as trying to study the evolution of
extremely blue and red objects. The color information used for sample
selection was thus deemed to be quite important.  For this reason, we elected
to use aperture colors, and magnitudes, when creating the spectral sample.

\begin{figure} [htbp]
  \begin{center}\epsfig{file=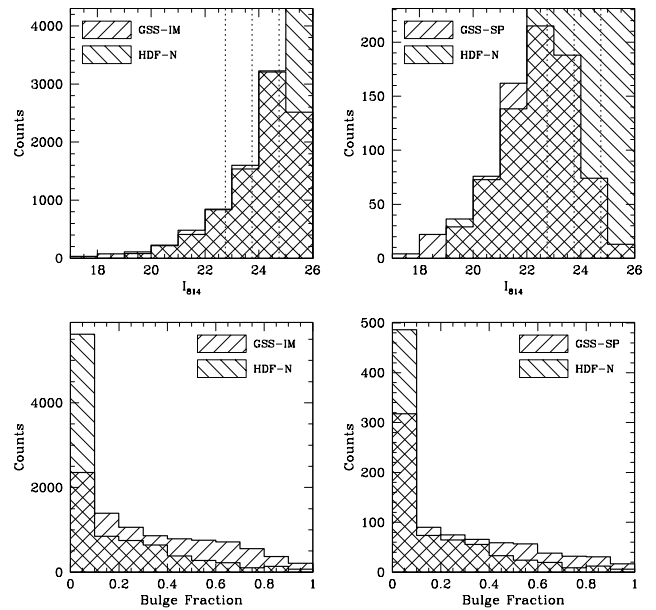, width=3.3in}\end{center}
  \caption[Figure 5]
{The distribution of number counts and bulge fractions as a function of
$I_{814}AB$ magnitude.  Dotted lines on the top panels show the three bins
used in the spectroscopic target selection, with a fourth bin for fainter
objects extending down to $I_{814}AB$ = 26.  The top left hand panel shows the
distribution of number counts for the 9212 objects in the GSS imaging catalog
and the 522 within the HDF-N field imaging catalog (Marleau \& Simard 1998).
The HDF-N catalog extends considerably deeper at faint magnitudes ($I_{814}AB
> 25$) than the GSS catalog, due to the extremely deep, repeated exposures of
the field.  However, there is quite good agreement in the distribution within
the limits of our spectral sample ($V_{606} + I_{814})/2 \le 25$ (note that
($V_{606} + I_{814})/2$ in the Vega system is roughly equivalent to $I_{814}$
in the AB system).
The top right hand panel compares the number counts for the GSS spectral
sample with the HDF-N field catalog.  There is reasonable agreement in the
distribution within the limits of the brightest bin of our spectral sample
where ($V_{606}$ + $I_{814})/2 \le 23$.
The bottom panels compare the bulge fraction distribution within the HDF-N
imaging catalog and the GSS spectral sample, finding an excess of no bulge
component objects in the HDF-N sample which is concentrated at faint
magnitudes and does not hold for objects brighter than $I_{814}AB = 23$.}
  \label{fig05}
\end{figure}

Figure~\ref{fig05} evaluates the depth of the HST+WFPC2 imaging, and the
consequent limits to the photometric catalogs, by comparing two analyses of
$I_{814}$ data conducted with the GIM2D package.  We compare the distribution
of number counts (Marleau \& Simard 1998) within the drizzled and deeply
imaged HDF-N field (Williams \etal 1996) with those within the GSS imaging
catalog, scaling the absolute number of counts (522 versus 9212, for objects
brighter than $I_{814} = 26$) by the relative surface areas (4.5 versus 127
square arcminutes), and find good agreement down to $I_{814} = 25$, with major
differences becoming apparent beyond 25 magnitudes.  Note that $V_{606}$ data
were not used in the GIM2D HDF-N analysis and are thus not available for
comparison, and also that the HDF-N I-band magnitudes were cataloged under the
AB system.  A parallel comparison with the 785 objects in the GSS spectral
sample with GIM2D parameters yields more scatter, but the agreement is still
reasonable for objects brighter than $(V_{606} + I_{814})/2 \le 23$.

The distribution of bulge fractions within the HDF-N sample contains an
increase in objects best fit with no bulge component, relative to both the GSS
imaging and the GSS spectral samples.  This effect is strongest for faint
objects, and is not significant below 23 magnitudes.  This is both because the
difference in bulge fraction distributions is not present in the brighter
objects, and because the number of objects has dropped to 72 for HDF-N sample,
decreasing the statistical significance of the comparison process.  Given the
catalog limits, we speculate that these objects may be low surface brightness
late-type spiral galaxies, falling below the surface brightness sensitivity of
the GSS samples.

In summary, we estimate the $5 \sigma$ detection limit of the imaging catalogs
to lie at $(V_{606} + I_{814})/2 = 25$ (though note that surface brightness
completion becomes an issue for significantly brighter objects, at roughly 23
magnitudes).  Note that errors on individual magnitudes are 0.02 for observed
quantities, and 0.02 -- 0.03 for derived rest-frame magnitudes (see Simard
\etal\ 2002 for complete details of such).

\section{Sample Selection}

A series of spectroscopic samples was then selected, using the Core
($1.5^{\prime\prime}$ diameter aperture) magnitudes and the Core
($1.0^{\prime\prime}$ diameter aperture) colors to define the distribution of
fluxes.  Targets were selected to populate three magnitude bins, and divided
into five color bins as well, for a general redshift survey.
Table~\ref{tab02} shows the distribution of the completeness ratio (fraction
of objects from the imaging catalog within each bin targeted for spectroscopy)
for the total spectral survey program, which includes a set of objects chosen
for additional properties (described in detail below) as well as the general
redshift survey.  

The primary magnitude limit for the general survey was $(V_{606} + I_{814})/2
= 24$, well above the detection limit of 25 for photometric completeness
(though see Simard \etal\ 1999 for a discussion of the effect of surface
brightness biases).  This limit was chosen to lie well above the completeness
level of the optical catalogs, as established by comparison with the data for
the deeply imaged HDF-N field (see Figure~\ref{fig05}), and to obtain
sufficient spectral flux for redshift determination of objects exhibiting a
range of morphologies (\eg both absorption and emission dominated spectra).

There are 83 objects with magnitudes between 24 and 25 for which spectra were
obtained, for the following reasons.  The pool of viable targets was
deliberately widened to 25 magnitudes when sampling chevron 7, to extend the
spectral survey to fainter magnitudes in the region where our photometry was
significantly deeper than the norm.  This resulted in observations of 28
objects which lay below the primary magnitude limit.  Six additional faint
objects were included as they were high priority targets for special
subsamples (R\#, hi-z, or phz, defined in Table~\ref{tab03}).  Twelve more
``filler'' objects were included due to a lack of brighter previously
undetected spectral candidates within a given ``slit position" along the GSS,
serving as candidates of last resort for the north-eastern end of the GSS
where our spectral observations were most concentrated.  Finally, 37 objects
with magnitudes between 24 and 25 were obtained serendipitously, falling
within slits designed for brighter targets.  There are also six fainter {\it
serendipitous} objects within the spectral sample, lying between 25$^{th}$ and
25.5$^{th}$ magnitude.

Note that Table~\ref{tab02} contains data on 788 objects from the
spectroscopic sample, omitting 30 objects for which Core magnitudes could not
be measured due to object close proximity to a WFPC2 chip edge or interfering
stellar diffraction spikes.  In addition, five stars selected for use in the
mask alignment process are contained within this sample that were not listed
in the spectroscopic sample described in Weiner \etal\ 2005a.

\begin{table*} [htbp]
  \caption{Observed Sample Completeness}
  \small
  \begin{center}
  \begin{tabular} {r c c c r l r r r r} 
  \hline
  \hline
  \multicolumn{6}{c}{Color} &  \multicolumn{4}{c}{$m = (V+I)/2$}  \\
                    & & & & & & $m \le 23$ & $23 < m \le 24$ & $24 < m \le 25^a$ & $25 < m \le 25.5^b$ \\
  \hline
           &   & $V-I$ &$\le$& 0.25 & (bluest)   &   1/2   &    0/2    &    1/99   & 1/194  \\
      0.25 &$<$& $V-I$ &$\le$& 0.50 & (blue)     &   6/16  &   29/90   &   10/454  & 2/387  \\
      0.50 &$<$& $V-I$ &$\le$& 1.75 &            & 333/941 &  229/1354 &   60/2613 & 3/1372 \\
      1.75 &$<$& $V-I$ &$\le$& 2.00 & (red)      &  36/85  &   27/72   &    2/52   & 0/19   \\
      2.00 &$<$& $V-I$ &     &      & (reddest)  &  30/61  &   15/55   &    3/56   & 0/28   \\
  \hline
  \hline \\ [-0.1in]
  \multispan{10} $^a$ The bulk of the objects within one magnitude below the spectroscopic survey limit are split            \hidewidth \\
  \multispan{10} between {\it serendipitously} observed objects, which fell within slits designed for brighter targets, and  \hidewidth \\
  \multispan{10} objects drawn from the single deeply imaged WFPC chevron number 7.                                          \hidewidth \\
  \multispan{10} $^b$ All six objects fainter than 25$^{th}$ magnitude are {\it serendipitous} detections.                   \hidewidth \\
  \end{tabular}
  \end{center}
  \label{tab02}
\end{table*}

Figure~\ref{fig06} shows the distribution of the imaging and the spectral
samples in color and in magnitude.  We prioritized objects with blue colors
($V_{606} - I_{814} < 0.5$) or with red colors ($V_{606} - I_{814} > 1.75$) in
the selection process; they are thus over-represented in the spectral sample
at a higher fraction (150\%) than objects with intermediate colors in the
range above 24$^{th}$ magnitude.  Because the five color bins were chosen to
be symmetric about the total color distribution, however, this did not
significantly affect the relationship of average color as a function of
magnitude in the spectral subsample.  In both samples, average $V_{606}$ -
$I_{814}$ colors are relatively constant for magnitudes brighter than 23 and
then begin to shift blueward, at a rate of 0.2 magnitudes in color per unit of
total magnitude.

\begin{figure} [htbp]
  \begin{center}\epsfig{file=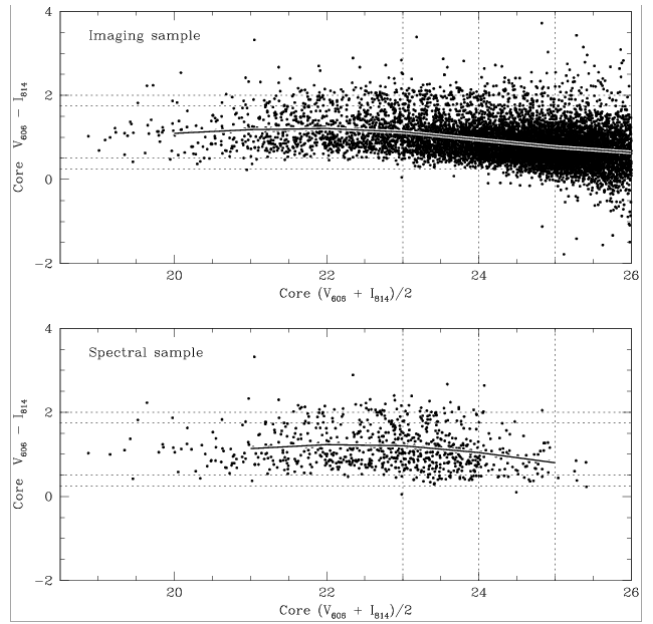, width=3.3in}\end{center}
  \caption[Figure 6]  
{The distribution of Core color ($1.0^{\prime\prime}$ diameter aperture)
versus Core magnitude ($1.5^{\prime\prime}$ diameter aperture) for the
magnitudes used in spectral target selection, for (a) the complete imaging
catalog and (b) the spectral subsample selected to observe with Keck+LRIS.
Colors have been averaged in half-magnitude running bins, and the average
values are connected with a line against the field of objects to show the mean
trend.  The dotted lines divide the space into fifteen blocks brighter than
$(I_{814} + V_{606})/2 = 25$; each bin was tracked separately through the
spectral target selection process.}
  \label{fig06}
\end{figure}

\section{Special Subsamples}

One could say that the spectroscopic program was designed in an umbrella
factory, in the sense that a host of scientific subprograms motivated the
inclusion of various subsamples of objects, chosen with additional criteria
beyond the described optical magnitudes and colors.
In addition to the general magnitude limited sample defined above, several
types of objects were prioritized in small numbers for selection from the
photometric catalog.  
Table~\ref{tab03} gives a brief description of these additional selection
criteria and their fractional representation within the complete spectral
sample.  We emphasize that almost all of these objects were also eligible for
selection as a part of the general redshift survey (the exception being six
serendipitously observed extremely faint objects which fall beyond the
magnitude limit).

\begin{table} [htbp]
  \caption{Selection Criteria Keys}
  \begin{center}
  \begin{tabular} {l l r r} 
  \hline
  \hline
  Key        & Description & \multicolumn{1}{c}{\#} & \multicolumn{1}{c}{\%} \\
  \hline 
  zsurvey    & redshift sample                                 & 566 &  69.2 \\ 
  serendip   & companions (on shared slitlet)                  & 121 &  14.8 \\ 
  disk	   & elongated galaxies, I $<$ 22.5                  &  64 &   7.8 \\ 
  align      & mask alignment pointers (mostly stars)          &  27 &   3.3 \\ 
  phz        & photometric redshift calibrators                &  12 &   1.5 \\ 
  R\#        & optical counterparts to known radio source      &   9 &   1.1 \\ 
  morph	   & morphologically selected                        &   7 &   0.9 \\ 
  group A, B & members of 2 small groups on the sky            &   6 &   0.7 \\ 
  hi-z 	   & dropout candidates                              &   5 &   0.6 \\ 
  cfrs	   & CFRS targets with no redshifts                  &   1 &   0.1 \\ 
  total	   & all spectral targets                            & 818 & 100.0 \\ 
  \hline \\ 
  \end{tabular}
  \end{center}
  \label{tab03}
\end{table}

The bulk of the spectral targets were selected for the general redshift survey
program (keyword {\it zsurvey}).
In addition to these primary targets, which were each placed within an
individual slitlet, we acquired spectra for a set of secondary objects ({\it
serendip}), which also fell within the same slitlets.  Thus, multiple objects
were on occasion observed within a single slitlet.  This usually occurred by
chance, though occasionally by design, where two objects fell together within
a single default slit length, along or near to the default mask angle
(40.5$^{\circ}$).  Some of the serendipitously detected objects are extremely
faint, and thus we would never have selected them as primary candidates for
spectroscopic follow-up due to the difficulty in getting enough spectral flux
into the slitlet.  These objects (characterized by faint broad band magnitudes
and very low spectral continuum levels) were typically detected through the
fortuitous discovery of an extremely bright, isolated emission line.

A third set of objects ({\it disk}) were chosen to allow the acquisition of
spatially resolved velocity profiles (Vogt \etal\ 1996, 1997, 2005).  In spite
of the misleading keyword, these objects were not selected for a disklike
morphology.  They were chosen instead for angular elongation on the sky,
equivalent to an inclination angle greater than 30$^{\circ}$ for a late-type
spiral disk, and for a position angle such that the major axis lay within
30$^{\circ}$ of the Keck+LRIS mask angle (held at 40.5$^{\circ}$).  These
simple criteria allowed the selection of early type galaxies, for example (\ie
for which a spatially extended spectrum, composed purely of continuum and
absorption features but with no emission was typically observed).  Because of
the need for high S/N spectra, these objects were prioritized only at the
bright end of the luminosity distribution (I$_{814} \le 22.5$).

Every Keck+LRIS mask which was constructed (see description below) contained a
set of slitlets for science objects, and then because the targets were
uniformly faint a small number of additional boxes (from three to six square
slitlets) were added for bright, pointlike objects ({\it align}) used in the
mask alignment process.  All but two of these objects were stars (plus one QSO
and one compact galaxy), and as these alignment objects were observed along
with the science targets we have spectra for them (albeit at lower resolution,
as the boxes are two arcseconds wide).

As mentioned earlier, we prioritized a set of faint, extremely blue targets
({\it phz}) because we lacked spectra to match HST-resolution images of such
targets in sufficient numbers to calibrate photometric redshift models.

The remaining four subsamples each contain less than 10 members.  For the
first set of spectroscopy observations of the leftmost one-third of the strip
(see Figure~\ref{fig02}), we overlaid the detections found in the field at
radio wavelengths within chevrons 6 or 8-10 (Fomalont \etal\ 1991) and placed
slitlets upon those for which an optical counterpart brighter than 24$^{th}$
magnitude existed ({\it R\#}).
An additional set of candidates was chosen, particularly in the deeply imaged
chevron 7, by eye for morphological peculiarities ({\it morph}).  They each
have either a diffuse light distribution, or appear to be a potential merger
in progress, or are a gravitational lens candidate (one object).
Six objects well within the magnitude limits of the redshift sample were
included because they appeared to make up two small groups in projection on
the sky ({\it group A, B}).
A final two objects were included because they fell within the region of the
strip which had been observed as a part of the CFRS ({\it cfrs}) redshift
survey (Lilly \etal\ 1995), and redshifts had not been determined from the
CFHT spectra.

Combining all objects placed within slitlets and the objects placed within
square boxes for alignment purposes, spectra were obtained for a total of 818
unique objects.

Due to the additional selection criteria, the total set of spectroscopic
targets is not a strictly random subsample of the photometric catalog as
defined by our magnitude limits.  In practice, however, the observed objects
sample the range of apparent color and magnitude in the photometric catalog
fairly evenly in the range $(V_{606} + I_{814})/2 \le 24$, as shown in
Figures~\ref{fig07} through \ref{fig11} for key structural parameters.  Note
that Simard \etal\ 2002 and C.~N.~A. Willmer \etal\ (2005, in preparation)
also discuss the detailed selection map in apparent color-magnitude bins.

\section{Modeling Sample Selection}

A set of Monte Carlo simulations was run, selecting objects from the imaging
catalog purely by aperture magnitudes and colors in the fifteen bins to match
the fractions listed in Table~\ref{tab02}.  We began by trimming the
spectroscopic sample from 818 to 782 objects.  We first removed six faint,
serendipitously observed objects which were fainter than the sample limit of
25 magnitudes.  We then removed 30 objects for which Core magnitudes could not
be measured, due to close proximity to WFPC2 chip edges or interfering stellar
diffraction spikes.  The imaging catalog was cut in a similar fashion to 6041
objects with both magnitudes brighter than 25 and a valid entry in the GIM2D
catalog. (Note that 947 objects brighter than 25 were fit by FOCAS but not by
GIM2D, which requires a larger background area around the object on the sky.)

Each ``mock spectroscopic catalog'' contains 782 objects, distributed
identically to the spectroscopic sample within the five color bands and three
magnitude bands used in sample creation.  Figure~\ref{fig07} shows the
distribution of magnitudes and colors for the spectroscopic sample and ten
mock samples, finding reasonable agreement given the scatter in the mock
catalog distributions.  The number of objects within each broad magnitude bin
is identical for the Core magnitudes and colors, by design.

We observe that the spectroscopic sample contains a slight excess of the
brightest objects within the 23$^{rd}$ magnitude band (and thus a dearth of
the faintest objects), due in part to the preferential inclusion of bright
stars used in the mask alignment process.  Within the 24$^{th}$ and 25$^{th}$
magnitude bands, we again see a slight bias towards including brighter objects
at the expense of fainter ones (\ie more objects between 23 and 23.5, or 24
and 24.5, than expected, coupled to less objects between 23.5 and 24, or 24.5
and 25).  This effect is not larger than the variance in magnitudes as
measured by different techniques, and becomes washed out in the right hand
panel for magnitudes derived with the GIM2D package.

\begin{figure} [htbp]
  \begin{center}\epsfig{file=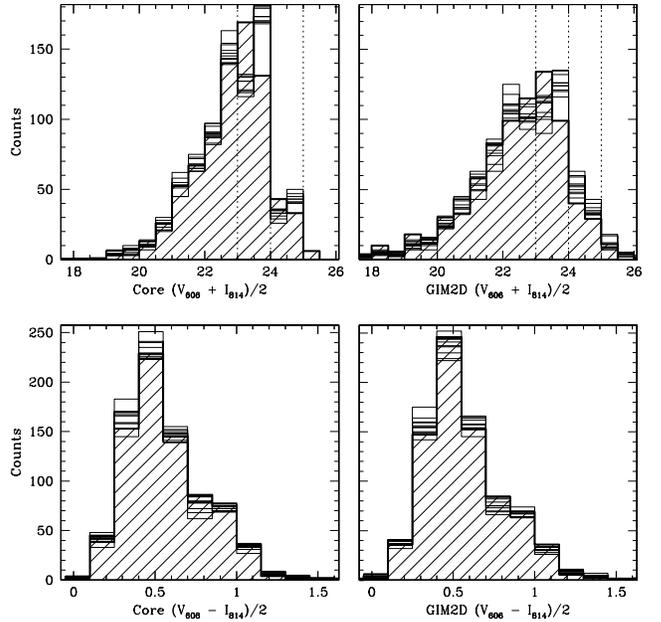, width=3.3in}\end{center}
  \caption[Figure 7] 
{The distribution of Core and GIM2D magnitudes and colors, for the
spectroscopic sample (the hatched histogram, outlined with a heavy line) and
ten simulated samples drawn from the imaging catalog with the magnitude and
color selection criteria for the spectroscopic survey but using no additional
criteria (\eg morphology).  
There is reasonable agreement between the simulated and the actual samples,
though a slight trend exists towards brighter objects within each broad
magnitude bin for the spectroscopic sample Core magnitudes (caused in part, in
the brightest magnitude bin, by the inclusion of bright stars used as mask
alignment objects.}
  \label{fig07}
\end{figure}

\begin{figure} [htbp]
  \begin{center}\epsfig{file=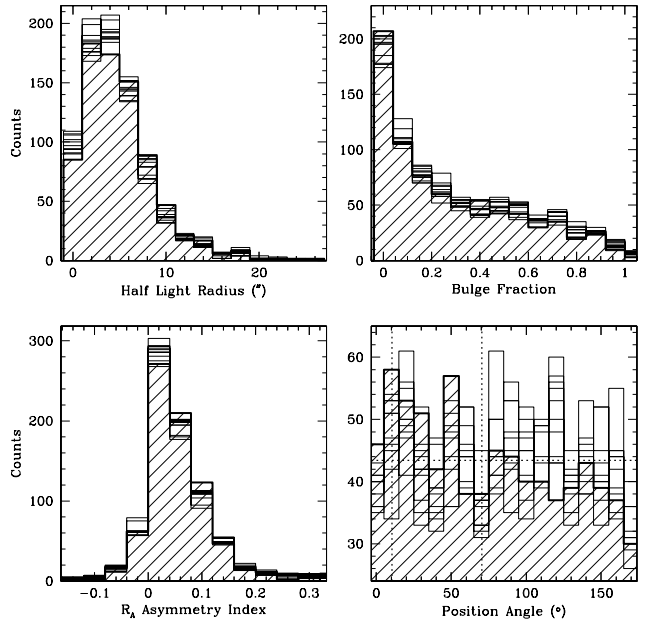, width=3.3in}\end{center}
  \caption[Figure 8] 
{The distribution of angular half-light radii, bulge fractions, $R_A$
asymmetry indices, and position angles on the sky, for the spectroscopic
sample (the hatched histogram, outlined with a heavy line) and ten simulated
samples drawn from the imaging catalog with the magnitude and color selection
criteria for the spectroscopic survey but using no additional criteria (\eg
morphology).  
There is reasonable agreement between the simulated and the actual samples for
the first three panels.  The $4\sigma$ bias in position angle towards low
values (the dotted horizontal line represents a perfectly even distribution)
is caused by the preferential selection of elongated objects for which the
position angle of the major axis lay within 30$^{\circ}$ of the Keck+LRIS mask
angle of 40.5$^{\circ}$ (the region marked by dotted vertical lines) within
the spectroscopic sample.  
}
  \label{fig08}
\end{figure}

Figure~\ref{fig08} shows the distribution of half light radii, bulge
fractions, $R_A$ asymmetry indices (also defined as $ra3$; \cf Schade \etal\
1995; Simard \etal\ 2002), a measure of the residual flux remaining for each
object after subtracting a doppelganger formed by rotating the object image
through 180$^{\circ}$ on the sky, and position angle on the sky for the
spectroscopic sample and ten mock samples.  There is a slight trend towards
including larger objects (by angular size) in the spectroscopic survey.
The one solid difference between the spectroscopic sample and the mock
samples, however, is found when examining the distribution of position angles
on the sky.  This distribution should be relatively flat, though the variance
in the mock sample distribution indicates the level of scatter.  As expected,
we observe an excess of objects within the spectral sample at low position
angles.  The cause for this trend is the prioritized selection of elongated
objects ({\it disk} classification) for which the position angles lie within
30$^{\circ}$ of the angle on the sky at which the Keck+LRIS masks were placed,
causing an excess of objects within 10$^{\circ}$ and 70$^{\circ}$.  There is a
12\% increase across this range, a $4\sigma$ variation given the scatter in
the mock catalog distributions.  

Note that this selection effect affects the position angle distribution but
does not translate into a significantly larger than expected fraction of, for
example, objects with small bulge fractions, as shown in the top panel.
Accounting for the 27 mask alignment objects (\eg excess stars, which tend to
be fit with a bulge profile of zero) within the spectral sample, no
significant differences remain in bulge distributions.
In summary, the selection of a small subset of objects based on additional
factors (\eg apparent morphology) in the actual spectral sample does not
appear to have significantly affected the distribution of structural
parameters other than position angle on the sky.

We examine the distribution of objects within the complete imaging catalog in
Figure~\ref{fig09}, plotting half light radii, bulge fractions, and $R_A$
asymmetry indices against magnitude.  In order to characterize the
distributions, the data were binned into a $20 \times 20$ grid within each
panel.  Contours were then fit to each grid image, running from 5\% to 75\% of
the peak values.

Figure~\ref{fig10} contains the subset of data forming the spectral sample,
plotted in similar fashion.  However, the contours in Figure~\ref{fig10} are
not fit to the spectral sample.  Instead, the grid images for the total
imaging catalog from Figure~\ref{fig09} were convolved with a spectral
luminosity selection function, and used to form a set of contours which
reflected the magnitude selection of the spectral sample.  This selection
function was formed by sampling the ratio of counts within the spectral sample
to counts within the parent imaging catalog, for each of the 20 bins formed by
the columns of the contour image grid (analogous to dividing the counts in the
first panel of Figure~\ref{fig10} by those in the first panel of
Figure~\ref{fig09}).  The effect is to multiply each grid point by the
fraction of objects selected for spectroscopic follow-up within the associated
luminosity bin.  

The resultant contours match the distribution of the spectra sample quite
well, and are in fact indistinguishable from contours drawn directly from the
spectral sample data.  Figure~\ref{fig11} shows this comparison explicitly,
overlaying the two sets of contours against each other.  The differences
between the distributions within each panel are not statistically significant.

\begin{figure} [htbp]
  \begin{center}\epsfig{file=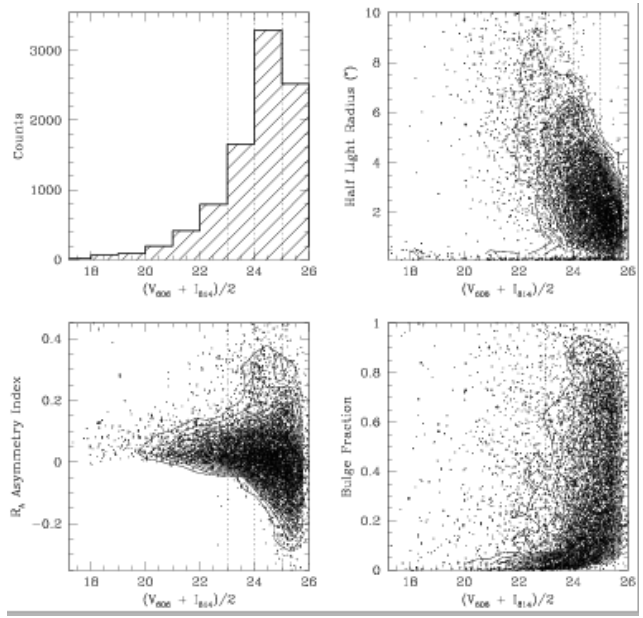, width=3.3in}\end{center}
  \caption[Figure 9] 
{The distribution of (a) number counts, (b) half-light radii, (c) $R_A$
asymmetry indices, and (d) bulge fractions as a function of magnitude for all
objects within the imaging catalogs, as determined with the GIM2D structural
analysis package.  Dotted lines on every panel separate the points into the
three bins used in the spectroscopic target selection, with a fourth bin for
fainter objects extending down to ($V_{606}$ + $I_{814}$)/2 = 26.  The twelve
contours overlaid on the last three panels extend from 5\% to 75\% of peak
values.}
  \label{fig09}
\end{figure}

\begin{figure} [htbp]
  \begin{center}\epsfig{file=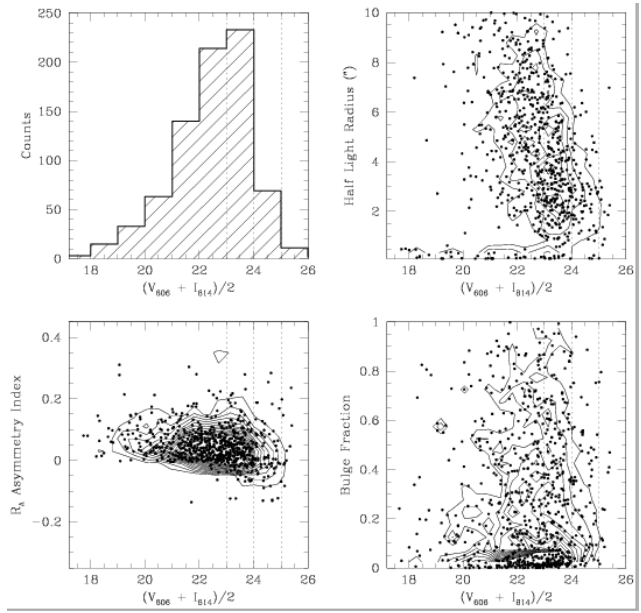, width=3.3in}\end{center}
  \caption[Figure 10]  
{The distribution of (a) number counts, (b) half-light radii, (c) $R_A$
asymmetry indices, and (d) bulge fractions as a function of magnitude for the
spectroscopic sample.  Dotted lines on every panel separate the points into
the three bins used in the spectroscopic target selection, with a fourth bin
for fainter objects extending down to ($V_{606}$ + $I_{814}$)/2 = 26.  The
spectral sample is biased toward the bright end of the magnitude selection
bins; the fraction of imaging targets contained within the spectral sample
drops off significantly below $V_{606}$ + I$_{814} = 24$, in the two faintest
bins (due to the increase in the number of possible targets and the decrease
in targeting priority below this limit).  Note that a comparison of bulge
fraction histograms between the imaging and spectroscopic samples, showing
good agreement, can be found by comparing the two lower panels in
Figure~\ref{fig05}.
The twelve contours overlaid on the last three panels were created by
convolving the distribution of the parent imaging sample, shown in
Figure~\ref{fig09}, with the fraction of objects selected as a pure function
of luminosity -- they are not fit to these data.  Note the agreement between
the distribution of the spectral sample and the contours, indicating that our
complex selection function can be treated as one of magnitude without bias in
these key parameters.}
  \label{fig10}
\end{figure}

\begin{figure*} [htbp]
  \begin{center}\epsfig{file=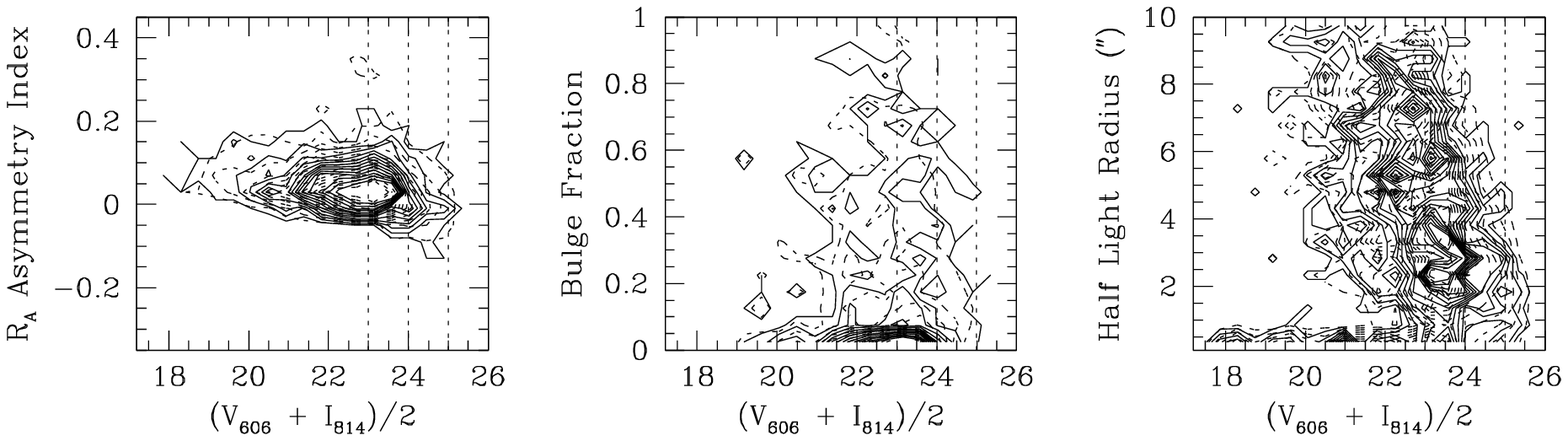, width=6.5in}\end{center}
  \caption[Figure 11]  
{A set of contours fit to the distribution of the spectral sample (solid
lines) are compared to those from Figure~\ref{fig10} (dotted lines), which
were creating by convolving a set of contours fit to the distribution of the
total imaging catalog with the fraction of objects observed in the spectral
sample as a function of magnitude.  In the first panel, note that the small
dotted contour in the region of high $R_A$ values falls at the lowest contour
(5\% of peak values), and is not statistically significant.  Both
distributions of bulge fractions (second panel) show a strong peak for objects
with extremely low bulge fractions, while the distributions of half light
radii (third panel) are broader.  There is good agreement between the two fits
in every panel.}
  \label{fig11}
\end{figure*}

\section{Mask design}

The DEEP1 spectra were taken as part of a multi-year spring season
spectroscopy program at the Keck Observatory, using the LRIS spectrograph (Oke
\etal\ 1995).  Observations of the GSS were conducted over nine observing runs
on 25 nights within a five-year period, in conjunction with several other
spring fields, and thus the process and technique for object selection was
adapted with passing years as the technology of manipulating large field
images matured (\eg memory allocation).  A total of 36 slit masks was
constructed for the strip, each containing between 30 and 55 objects.

An initial set of fifteen LRIS slit masks was constructed in 1995.  We had no
equipment which would allow us to create slitlets at an arbitrary mask angle
at this time, so spectra were taken for all objects at the same position angle
on the sky (40.5$^{\circ}$) regardless of the angle of any spatial elongation
of the object.  Individual slitlets were 1$^{\prime\prime}$ in width and
12$^{\prime\prime}$ in length, and each mask contained roughly 30 of them.
The bulk of the objects was selected for the general redshift survey, and a
set of alignment objects was included on every mask.

The Groth Strip is roughly four arcminutes wide, and so we attempted to place
objects from only the upper or the lower half within a single mask, so as to
keep the shift in observed spectral range as small as possible from object to
object.
Once objects were selected for the spectral program, they were placed upon
either one, two, or four masks, depending upon the faintness of the target.
Objects brighter than 23 in ($V_{606}$ + $I_{814}$)/2 were placed initially on
a single mask, those between 23$^{rd}$ and 24$^{th}$ magnitude were placed on
two masks, and the faintest objects, below 24$^{th}$ magnitude (drawn
primarily from the deeply imaged chevron 7), were placed on four masks to
maximize the amount of spectral flux that we would obtain.

Throughout the spectroscopy program, each mask was observed in turn with a
blue wavelength grating (typically 900~l~mm$^{-1}$/5500\AA) and a red
wavelength grating (600~l~mm$^{-1}$/7500\AA), to create a combined spectrum
covering the range 5000\AA\ to 8200\AA.  We planned for $2 \times 1500$
seconds exposure per grating per mask.

From 1996 onward, the minimum acceptable slit-length was allowed to drop to
8$^{\prime\prime}$, as one could extract usable spectra reliably with
background regions on each side of the object of smaller spatial extent due to
the overall brightness of the continuum level and strong key emission and
absorptions features. A small fraction of all slits were tilted at an angle
(between $10.5^{\circ}$ and $70.5^{\circ}$ eastwards of north), in order to
trace along the major axis of a spatially elongated object or to capture two
objects within one slit.  As a result of these changes, the average number of
objects per mask rose to 45.

The effect of the distribution on masks along and across the entire Groth
Strip can be observed in the lower panels of Figure~\ref{fig02}. The large,
broad peak at -20 arcminutes in the left hand panel reflects our initial focus
in a region surrounding the deeply imaged chevron 7.  Note that the spectral
fraction, the fraction of the imaging catalog objects which were observed with
Keck+LRIS, occasionally rises above a level of one, as all spectral targets
were counted but the normalization was to the limits of the general redshift
survey (($V_{606}$ + $I_{814}$)/2 < 24).  The right hand panel shows a slight
increase in spectral fraction along the top half of the chevrons (chips 2 and
3), caused by the distribution of masks along either the top of the bottom of
the strip.

\begin{figure*} [htbp]
  \begin{center}\epsfig{file=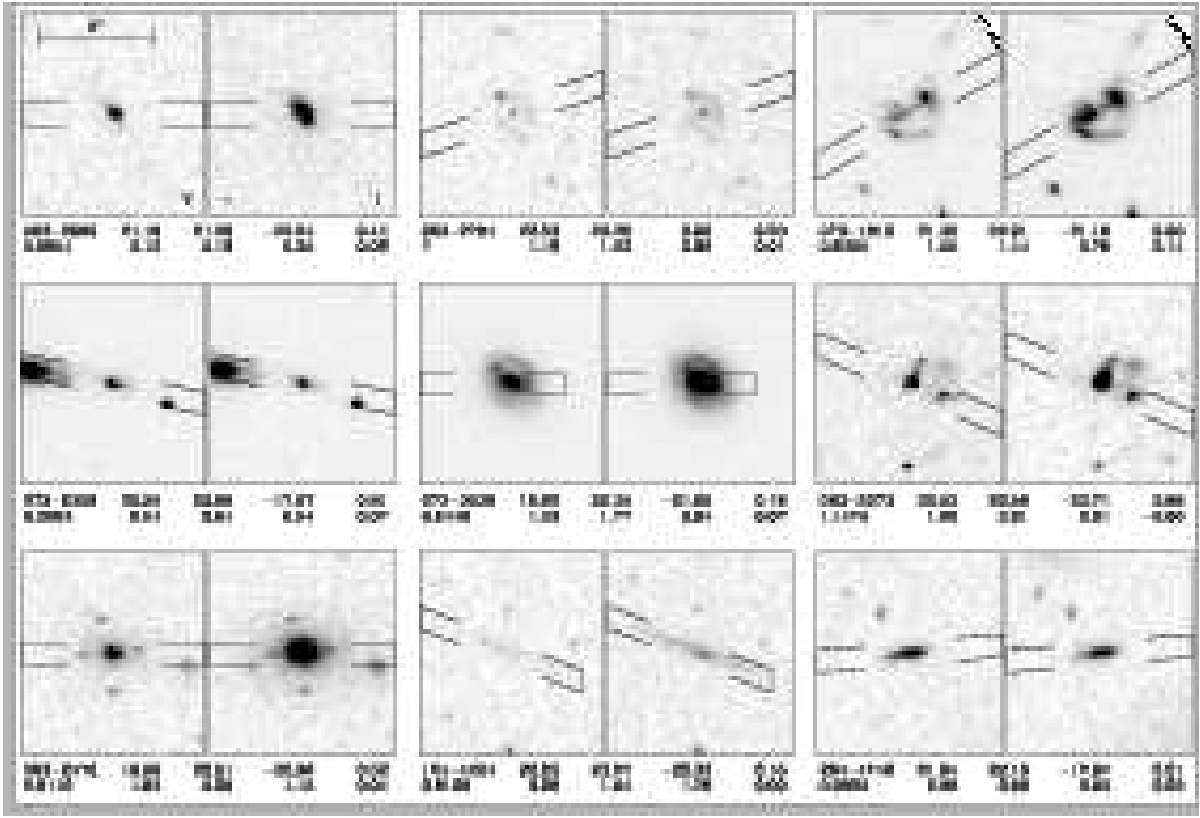, width=0.60\linewidth, angle=-90, clip=}\end{center}
  \caption[Figure 12] 
{A representative set of $V_{606}$ and $I_{814}$ HST+WFPC2 images of objects
within the spectral sample.  The solid lines on each postage stamp image
indicate the position of the KECK+LRIS slitlet upon the object.  Each object
is labeled by name, followed by total $I_{814}$ GIM2D magnitude, aperture
$I_{814}$ Core magnitude, absolute B-band magnitude, and bulge fraction; the
lower line contains redshift, $V_{606} - I_{814}$ GIM2D color, aperture
$V_{606} - I_{814}$ Core color, half light radius in arcseconds, and asymmetry
parameter $R_A$.}
  \label{fig12}
\end{figure*}

\section{Postage stamps} 

A catalog of annotated $V_{606}$ and $I_{814}$ images has been created for the
entire spectral sample observed with Keck+LRIS (Figure~\ref{fig12} is a
representative sample).  Each image is overlaid with the position of the LRIS
slitlet at which it was observed, and labeled with a set of spectral- and
image-derived parameters.  A single representative panel is shown here, and
the entire set is available at the DEEP web site at {\tt
http://deep.ucolick.org}.

\section{Summary} 

This paper describes the framework and observational basis for first five-year
phase (DEEP1) of the Deep Extragalactic Exploratory Probe (DEEP), a longterm,
multifaceted study of the formation and evolution of galaxies in the distant
Universe.  We have obtained complete HST imaging in $I_{814}$ and $V_{606}$ of
the Groth Survey Strip, a 127 square arcminute region in the northern
hemisphere sky, and conducted Keck+LRIS spectroscopy of 818 objects in a
redshift survey which extends down to magnitudes of $(V_{606} + I_{814})/2 =
25$.  

Though our selection criteria for the spectroscopic subsample are complex, and
have varied as the survey progressed, the bulk of the objects were selected to
match the range in colors and magnitudes at the bright end (\ie $(V_{606} +
I_{814})/2 \le 25$) of the imaging catalog.  We model the complete
spectroscopic subsample, and show that it can be reproduced statistically by
making straightforward cuts in magnitude and color in the parent imaging
catalog.  The sole exception to this statement lies in the distribution of
position angles on the sky, where we observe an expected bias towards objects
with major axes oriented along the position angle at which the Keck+LRIS masks
were aligned.

\acknowledgments

The authors recognize and acknowledge the cultural role and reverence that the
summit of Mauna Kea has always had within the indigenous Hawaiian community.
We are most fortunate to have the opportunity to conduct observations from
this mountain.
The authors thank the staffs of HST and Keck for their help in acquiring these
data, the W. M.  Keck Foundation and NASA for construction of the Keck
telescopes, and Bev Oke and Judy Cohen for their work on LRIS that enabled the
spectroscopic observations described herein.
We give our thanks to the anonymous referee of this manuscript, who showed a
careful and courteous eye for detail.
NPV is a Guest User, Canadian Astronomy Data Center, which is operated by the
Dominion Astrophysical Observatory for the National Research Council of
Canada's Herzberg Institute of Astrophysics.

This work was conducted under the auspices of the DEEP (Deep Extragalactic
Evolutionary Probe) project, which was established through the Center for
Particle Astrophysics.  Funding was provided by NSF grants AST-9529098,
AST-0071198, by AST-0349155 to NPV through the Career Awards program, and by
NSF--0123690 via the ADVANCE-IT Program at NMSU, and by NASA grants
GO-07883.01-96A, GO-10249.01A, and AR-05801.01-94A, AR-06402.01-95A, and 
AR-07532.01-96A.
HST imaging of the Groth Strip was planned, executed, and analyzed by EJG and
JR with support from NASA grants NAS5-1661 and NAG5-6279 from the WFPC1 IDT.
SMF would like to thank the California Association for Research in Astronomy
for a generous research grant.

\appendix
\section{APPENDIX A: Particulars of DEEP1 GSS Mask Design}

The DEEP1 Groth Strip spectroscopic survey was a multi-year pilot program,
designed to find the observational limits of the Keck+LRIS equipment and to
explore a variety of scientific programs.  For these reasons, as well as the
rapid parallel advances in computer technologies, the object selection process
evolved with time.  We present here a detailed overview of the creation
process for the entire set of 36 masks.

Sample selection for the first 15 masks was conducted by using a printed set
of black and white $I_{814}$ images of the target chevrons, mosaicked by hand
into a single, extended map.  Transparent overlays were coded to indicate
object Core magnitudes and colors.  
This first set of masks was placed to overlap with the deeply imaged chevron 7
at the northeast end of the strip (fourth chevron from the left end, in
Figure~\ref{fig02}), and extended from chevron 6 down to chevron 10.  The
first group of four masks (GSS1A, GSS1B, GSS1C, and GSS1D) had a single
position on the sky and contained objects selected from chips 2 and 3 of the
targeted chevrons, while the second and third sets of four (GSS2-ABCD,
GSS3-ABCD) contained objects selected from chips 3 and 4 and were offset
toward the lower edge of the Groth Strip.  The final three masks (GSSB-abc)
were back-up program masks designed for use in poor weather (low transparency
conditions), with 1.2$^{\prime\prime}$ width slitlets.  Each mask contained
the brightest candidates from the GSS1, GSS2, and GSS3 series respectively,
and were filled with additional bright objects in the range ($V_{606}$ +
$I_{814}$)/2 $\le 23$.

A few serendipitous spectra would be discovered during the spectral reduction
process, and the entire set of {\it radio} and {\it group} objects was
selected at this time, as well as several {\it morphological} candidates.
Note that no objects were selected according to the {\it disk}, {\it cfrs}, or
{\it phz} key words for these early mask sequences (though the bluest and
reddest objects were prioritized within the general redshift survey).

Beginning in 1996, masks were designed using interactive, computerized
selection to select objects and to place them on the masks (NPV's {\tt
disktool} IRAF package).  Following target and alignment star selection, all
masks were modeled with ACP's {\tt ucsclris} IRAF package to account for
specific observational constraints (\eg anamorphic corrections) and to create
machine (CAD) instructions for milling the mask locally in the UCSC
instrumentation shops.
The pattern of placing objects on one, two, or four masks according to
brightness was maintained.  In addition, any objects which were re-observed
because of insufficient initial flux levels (\ie no redshift could be
assigned) were placed on as many masks as were available, and objects with the
reddest colors were automatically placed on at least two masks regardless of
brightness (spectral reductions proved to be difficult for these objects, due
to the presence of very little other than wide, diffuse absorption features).
After the first year a small fraction of all slits were tilted at an angle
($\theta < 30^{\circ}$) relative to 40.5$^{\circ}$, in order to trace along
the major axis of a spatially elongated object or to capture two objects
within one slit.  For the {\it disk} candidates, exposure time was doubled
from one mask to two to increase S/N levels for rotation curve extraction.

In 1996, the thirteen unused masks were updated with several elongated {\it
disk} candidates, several additional {\it morph} candidates, and a set of the
{\it phz} objects with extremely blue colors.  Because KPNO UBRI photometry
had been obtained for the northeast portion of the strip in the intervening
year, these data were used as well in the relative prioritization of the 17
{\it phz} candidates.  The masks were updated in place, with original names
retained.  (Only masks GSS1-A and GSS1-B were used in 1995, due to bad
weather.)
In 1997 we created a new set of five masks (GSS4-ABCDE), shifting our default
position along the top of the strip slightly to the right to include chevron
11.  We ceased placing general targets on the masks which were fainter than
($V_{606}$ + I$_{814} = 24$, as we found them to demonstrate unlikely success
for redshift identification.  Five UB dropout candidates (with photometric
redshifts $z > 2.5$) were identified from the KPNO UBRI data and placed on
four masks each.  In 1998 seventeen additional masks were created (GSS5-ABC,
GSS6-ABCDEF, and GSS7-ABCDEFGH).  We began to employ tilted slits not only to
trace the major axes of individual spatially extended objects but also to
place secondary, and occasionally tertiary, {\it serendip} objects which were
offset from primary targets by $\theta < 30^{\circ}$ from the mask angle, to
increase efficiency.

The GSS5 series was designed to re-observe objects from within chevrons 6
through 11 for which we still did not have redshifts, in spite of repeated
observations.  We prioritized objects in the following order: (1) previously
observed targets with I$_{814} \le 23$ and no secure redshift, (2) previously
observed sources with I$_{814} \le 24$ and no secure redshift, (3) previously
observed sources with I$_{814} > 25$ and no secure redshift, (4) high redshift
(U and B dropout) candidates determined from ground based UBRI data or from
existing GSS spectra which showed promise (\ie no features indicating lower
redshifts), and (5) CFRS survey objects without redshifts.
The GSS6 series was placed at the southwest end of the strip.  Two new
pointings were established, and at each one two normal masks (ABDE) and one
back-up mask (CF) with bright objects was created.  The {\it disk}, {\it phz}
and extremely red objects were prioritized in the sample selection process.
The GSS7 series was placed in the middle region of the strip, designed to
sample the bright end of the luminosity function and bridge the spatial gap
between the northeast and southwest sampled portions.  Four new pointings were
created, and a normal mask (ACEG) and a back-up mask (BDFH) with bright
objects were made for each one.  The {\it disk}, {\it phz} and extremely red
objects were again prioritized in the sample selection process; in addition as
there would be only one mask exposure per position under optimum conditions,
we limited the general redshift survey to ($V_{606}$ + I$_{814})/2 \le 23$.

In 1999 ten additional masks were created (GSS6-GHIJ and GSS7-IJKLMN).  They
were placed in sets of two at five of the six positions established in 1998
(the sixth pointing, home of masks GSS7-EF was not doubled because it had not
been used in 1998).

We note two factors of interest to anyone studying interactions, pairs of
objects, or large scale structure: (a) the repeated placement of fainter
objects on a series of masks all centered at the same location, conducted to
increase the total spectral flux and thus the probability of redshift
determination for such objects, and (b) the requirement of a few (four to six)
arcseconds of slit being available to each side of each object for background
light removal.  These both affect the object-to-object angular distribution
within the spectral sample on small size scales ($\theta \sim
10^{\prime\prime}$).  The first effect will further result in a sparser
sampling of near neighbors on the sky for the fainter objects within the
spectral sample.

\section{APPENDIX B: Conversion Factors between AB and Vega Magnitudes}

The convolution between filter responses and galaxy SEDs followed Fukugita
\etal\ (1995), by resampling filters and spectra to the same dispersion (1
\AA), using parabolic and linear interpolations respectively.  For the HST
$I_{814}$ and $V_{606}$ filters, efficiency curves for WFPC2 were downloaded
from the Space Telescope Science Institute web site and curves for the
different CCDs were averaged.  The Vega spectrum used to calculate the
conversion to AB magnitudes is the model atmosphere calculated by Kurucz,
distributed with the Bruzual \& Charlot (2003) galaxy evolution synthesis
package. The AB spectrum is simply a flat spectrum in F($\nu$) converted into
wavelength space (\eg Fukugita \etal\ 1995).

\begin{table*} [htbp]
  \caption{Conversion Factors between AB and Vega Magnitudes}
  \begin{center}
  \begin{tabular}{l l l}
  \hline
  \hline
  \multicolumn{3}{c}{Transformation Between AB and Vega Systems} \\
  \noalign{\smallskip} \hline \noalign{\smallskip}
 
  $V606_{AB}$  &   =  & $V606_{Vega}$         +  0.096 \\ 
  $I814_{AB}$  &   =  & $I814_{Vega}$         +  0.417 \\ 
  ($V606-I814)_{AB}$& =& ($V606-I814)_{Vega}$ -- 0.321 \\
  \noalign{\smallskip} \hline
  \end{tabular}
  \end{center}
  \label{tab05}
\end{table*}

\vfill\eject

\end{document}